\documentstyle[twoside,fleqn,espcrc2,epsfig]{article}

\def\Journal#1#2#3#4{{#1} {\bf #2}, #3 (#4)}

\def\NPB{{\em Nucl. Phys.} B}
\def\PLB{{\em Phys. Lett.}  B}
\def\PRL{\em Phys. Rev. Lett.}
\def\PRD{{\em Phys. Rev.} D}

\def\ZPA{{\em Z. Phys.} A}
\def\NPA{{\em Nucl. Phys.} A}

\def\nn{\nonumber}

\def\be{\begin{equation}}
\def\ee{\end{equation}}
\def\bea{\begin{eqnarray}}
\def\eea{\end{eqnarray}}

\def\gev{\,{\rm GeV}}
\renewcommand{\c}{\rm{c}}
\newcommand{\g}{\rm{g}}
\renewcommand{\d}{\rm{d}}
\renewcommand{\u}{\rm{u}}
\newcommand{\q}{\rm{q}}
\newcommand{\cbar}{\overline{\rm{c}}}
\newcommand{\qbar}{\overline{\rm{q}}}

\newcommand{\psla}{p\kern-1.0ex/}
\newcommand{\qsla}{q\kern-1.1ex/}
\newcommand{\esla}{\epsilon\kern-1.0ex/}
\newcommand{\LQCD}{\Lambda_{\rm{QCD}}}
\def\BbB{B\overline{B}}
\def\jp{J/\psi}
\def\mjp{M_{J/\psi}}
\newcommand{\mc}{m_{\c}}
\def\als{\alpha_s}
\def\su3{{\rm SU}(3)_{{\rm F}}}

\newcommand{\da}{{DA}}
\newcommand{\das}{{DAs}}

\newcommand{\AmS}{{\protect\the\textfont2
  A\kern-.1667em\lower.5ex\hbox{M}\kern-.125emS}}

\title{{\bf Exclusive charmonium decays} \hspace{5cm} WU B 97-25}

\author{P.\ Kroll\address{Fachbereich Physik, Universit\"at Wuppertal\\
        Gau\ss strasse 20, D-42097 Wuppertal, Germany}%
\thanks{Supported in part by the TMR network ERB 4061 Pl 95 0115
         }}
       
\begin{document}

\begin{abstract}
The role of power corrections and higher Fock-state contributions to 
exclusive charmonium decays will be discussed. It will be argued that 
the $\jp$ ($\psi'$) decays into baryon-antibaryon pairs are dominated by the 
valence Fock-state contributions. $P$-wave charmonium decays, on the 
other hand, receive strong contributions from the $\c\cbar \g$ Fock 
states since the valence Fock-state contributions are suppressed in 
these reactions. Numerical results for $\jp (\psi') \to \BbB$ 
decay widths will be also presented and compared to data.\\

\noindent
Contribution to the QCD 97 conference, Montpellier (1997) 
\end{abstract}

\maketitle

\section{INTRODUCTION}
Exclusive charmonium decays have been investigated within perturbative
QCD by many authors, e.g.\ \cite{dun80}. It has been argued
that the dominant dynamical mechanism is $\c\cbar$ annihilation into the
minimal number of gluons allowed by colour conservation and charge
conjugation, and subsequent creation of light quark-antiquark pairs
forming the final state hadrons ($\c\cbar \to n\g^* \to m(\q\qbar)$). The
dominance of annihilation through gluons is most strikingly reflected
in the narrow widths of charmonium decays into hadronic channels in a
mass region where strong decays typically have widths of hundreds of
MeV. Since the $\c$ and the $\cbar$ quarks only
annihilate if their mutual distance is less than about $1/\mc$ (where
$\mc$ is the $\c$-quark mass) and since the average virtuality of the
gluons is of the order of $1 - 2\, \gev^2$ one may indeed expect
perturbative QCD to be at work although corrections are presumably
substantial ($\mc$ is not very large).\\ 
In hard exclusive reactions higher Fock-state contributions are
usually suppressed by inverse powers of the hard scale, $Q$, appearing
in the process ($Q=2\mc$ for exclusive charmonium decays), as compared
to the valence Fock-state contribution. Hence, higher Fock-state 
contributions are expected to be negligible in most cases.  
Charmonium decays are particularly interesting because the 
valence Fock-state contributions are often suppressed for one or the other
reason. In such a case higher Fock-state contributions or other
peculiar contributions such as power corrections or small components
of the hadronic wave functions may become important. In the following 
I present a few examples of charmonium decays with suppressed valence 
Fock-state contributions:
\begin{enumerate}
\item {\it Hadronic helicity non-conserving processes} (e.g.\
  $\jp\to\rho\pi$, $\eta_{\c}\to \BbB$,
  $\chi_{\c0}\to\BbB$). The standard method of calculating the valence
  Fock-state contributions leads to vanishing helicity non-conserving
  amplitudes. There are many attempts to understand the helicity
  non-conserving processes (e.g.\ intrinsic charm of the $\rho$ meson
  \cite{bro97}; diquarks in baryons \cite{kro:93a}) but a satisfactory
  explanation of all these processes is still lacking.
\vspace*{-0.1cm}
\item {\it $G$ parity violating $\jp$ decays}
  (e.g.\ $\pi^+\pi^-$, $\omega\pi^0$, $\rho\eta$).
  These reactions can proceed through the electromagnetic elementary
  process $\c\cbar\to\gamma^*\to n(\q\qbar)$ and/or may receive
  contributions from the isospin violating part of QCD. 
  In general, these contributions are rather small and thus
  other contributions may play an important role here.
\vspace*{-0.3cm}
\item {\it Radiative $\jp$ decays into light pseudoscalar mesons}.
  The purely electromagnetic
  process $\c\cbar\to\gamma^*\to\gamma\q\qbar$, a contribution of order
  $\alpha^3$ being proportional to the time-like $\pi\gamma$
  transition form factor at $s=\mjp^2$, together with a power
  correction provided by the VDM contribution $\jp\to\rho\pi$, 
  $\rho\to\gamma$ \cite{CZ84}, leads to 
  $\Gamma(\jp\to\pi^0\gamma) = 2.86\,$ eV in agreement with experiment
  (3.43$\pm$1.2 eV \cite{PDG}). Similar estimates of the 
  $\eta\gamma$ and $\eta'\gamma$ widths fail. Agreement with experiment
  is here obtained from a twist-4 gluon component of the singlet-$\eta$ state
  \cite{nov80}, i.e.\ from a power correction. That gluon component can
  occur as a consequences of the $U(1)$ anomaly. 
\vspace*{-0.2cm}
\item {\it $\chi_{\c J}$ decays.} For the $\chi_{\c J}$ mesons the $\c\cbar$
  pair forms a colour-singlet $P$-wave in the valence Fock
  state (notation: $\c\cbar_{1}(^3P_J)$). The next-higher Fock state is a
  $\c\cbar \g$  $S$-wave where the quark-antiquark pair forms a
  $\c\cbar_{8} ({}^3S _1)$ state. For this reason the latter
  contribution is customarily referred to as the colour-octet
  contribution. The colour-singlet and octet
  contributions to the $\chi_{\c J}\to h\overline h$ decay amplitude
  behave as \cite{BKS}
  \be
  \label{dim}
  M_J^{(c)}\,\sim f_h^2\, f^{(c)}({}^3P_J)\, \mc^{-n_c}\, .
  \ee
  The singlet decay constant, $f^{(1)}(^3P_J)$, which represents the
  derivative of a two-particle (non-relativistic) coordinate space 
  wave function at the origin, and the octet decay constant, 
  $f^{(8)}(^3P_J)$, as a three-particle wave function at the origin, 
  are of the same dimension, namely GeV$^2$. Hence, $n_1=n_8$. In
  fact, $n_c=1\, +\, 2n_h$ where $n_h$ is the dimension of the light 
  hadron's decay constant, $f_h$. It is, therefore, unjustified to 
  neglect the colour-octet contributions in the $\chi_{\c J}$ decays.
\end{enumerate} 

\section{VELOCITY SCALING}
Recently, the importance of higher Fock states in understanding the
production and the {\it inclusive} decays of charmonium has been
pointed out \cite{BBL}. As has been shown the
long-distance matrix elements can there be organized into a hierarchy
according to their scaling with $v$, the typically velocity of the $\c$
quark in the charmonium. One may apply the velocity expansion to {\it
exclusive} charmonium decays as well \cite{BKS}. The Fock-state
expansions of the charmonia start as
\bea
\label{Fockexpansion}
  |\jp\rangle &=& O(1)\, |\c\cbar_{1}({}^3S_1)\rangle
                        + O({\it v})\, |\c\cbar_{8}({}^3P_J)\, \g \rangle\nn\\
              &&        + O(v^2)\, |\c\cbar_{8}({}^3S_1)\, \g\g \rangle
                        + O({\it v}^3)\, , \nn\\
  |\;\chi_{\c J}\rangle &=& O(1)\, |\c\cbar_{1}({}^3P_J)\rangle
                      + O({\it v})\, |\c\cbar_{8}({}^3S_1)\, \g \rangle\nn\\
                    &&  + O(v^2). 
\eea
Combining the fact that the hard scattering amplitude involving a
$P$-wave $\c\cbar$ pair is of order $v$, with the
Fock-state expansion (\ref{Fockexpansion}), one finds for the
amplitudes of $\chi_{\c J}$ decays into, say, a pair of 
pseudoscalar mesons the behaviour
\bea
\label{ampPP} 
M(\chi_{\c J}\to P \overline{P}) &\sim& a\als^2 v + b \als^2
                 \big (v\sqrt{\als^{{\scriptscriptstyle soft}}}\big )\nn\\
                     &+&O(v^2),
\eea
where $a$ and $b$ are constants and $\als^{{\scriptscriptstyle soft}}$
comes from the coupling of the Fock-state gluon to the hard process. 
For the decay reaction $\jp\to\BbB$, on the other hand, one has
\bea
\label{ampBB}
M(\jp\to\BbB)&\sim&a \als^3 + b \als^3
                     v\big (v\sqrt{\als^{{\scriptscriptstyle
                     soft}}}\big )\\
        &+&c \als^3\; v^2 \als^{{\scriptscriptstyle soft}} + O(v^3).\nn   
\eea
Thus, one sees that in the case of the $\chi_{\c J}$ the
colour-octet contributions are not suppressed by powers of either 
$v$ or $1/\mc$ as compared to the contributions from the valence Fock 
states \cite{BKS}. Hence, the colour-octet contributions have to be
included for a consistent analysis of $P$-wave charmonium decays. 
Indeed, as an explicite analysis reveals \cite{BKS}, only if both the
contributions are taken into account agreement between predictions 
and experiment is obtained for the $\chi_{\c J}\to P\overline{P}$ decay widths. 
For more details and numerical results for decay widths, see the talk 
by G.\ Schuler in these proceedings. The situation is different for 
$\jp$ decays into baryon-antibaryon pairs: Higher Fock state
contributions first start at $O(v^2)$, see (\ref{ampBB}). Moreover, 
there is no obvious enhancement of the corresponding hard scattering 
amplitudes, they appear with at least the same power of $\als$ as 
the valence Fock state contributions. Thus, despite of the fact that 
$\mc$ is not very large and $v$ not small ($v^2\simeq 0.3$), it seems 
reasonable to expect small higher Fock-state contributions to the 
baryonic decays of the $\jp$.\\ 
In the following sections I will report on an analysis of the processes
$\jp (\psi')\to \BbB$ performed with regard to these observations \cite{bol97}.

\section{THE MODIFIED PERTURBATIVE APPROACH}
Recently, a modified perturbative approach has been proposed
\cite{BLS} in which transverse degrees of freedom as well as Sudakov
suppressions are taken into account in contrast to the standard
approach of Brodsky and Lepage \cite{lep80}. The modified perturbative
approach possesses the advantage of strongly suppressed end-point
regions. In these regions perturbative QCD cannot be
applied. Moreover, the running of $\als$ and the evolution of the 
hadronic wave function can be taken into account properly.\\
Within the modified perturbative approach an amplitude for a ${}^{2S+1}L_J$
charmonium decay into two light hadrons, $h$ and $h'$, is written as a
convolution with respect to the usual momentum fractions, $x_i$,
$x_i'$ and the transverse separations scales, ${\bf b_i}$, 
${\bf b'_i}$, canonically conjugated to the transverse momenta, of the
light hadrons. Structurally, such an amplitude has the form 
\bea 
\label{structure}
\lefteqn{M^{(c)}({}^{2S+1}L_J\to h h')\,=\, f^{(c)}({}^{2S+1}L_J)}\nn\\ 
  &\times& \int [\d x][\d x'] \int \frac{[\d^2{\bf b}]}{(4\pi)^2}
                   \frac{[\d^2{\bf b}']}{(4\pi)^2}\\
        &\times& \hat{\Psi}_h^{*}(x,{\bf b})\, 
                                  \hat{\Psi}_{h'}^{*}(x',{\bf b}')\nn\\
        &\times& \hat T_H^{(c)}(x,x',{\bf b},{\bf b}',t)
              \exp{[-S(x,x',{\bf b},{\bf b}',t)]},\nn
\eea
where $x^{(}{}'{}^{)}$, ${\bf b}^{(}{}'{}^{)}$ stand for sets of momentum 
fractions and transverse separations characterizing the hadron $h^{(}{}'{}^)$.
Each Fock state (see (\ref{Fockexpansion})) provides such an
amplitude (marked by the upper index $c$)\footnote{ In higher
Fock-state contributions additional integration variables may
appear.}. $\hat{\Psi}_h$ denotes the
transverse configuration space (light-cone) wave function of a light 
hadron. The $f^{(c)}$ is the charmonium decay constant already 
introduced in Sect.\ 1. Because the annihilation radius is
substantially smaller than the charmonium radius this is,
to a reasonable approximation, the only information on the charmonium 
wave function required. $\hat{T}_H^{(c)}$ is the Fourier transform 
of the hard scattering amplitude to be calculated from a set of
Feynman graphs relevant to the considered process. 
$t$ represents a set of renormalization scales at which the 
$\als$ appearing in $\hat{T}_H^{(c)}$, are to be evaluated. The 
$t_i$ are chosen as the maximum scale of either the longitudinal
momentum or the inverse transverse separation associated with each of
the internal gluons. 
Finally, $\exp{[-S]}$ represents the Sudakov factor which takes into
account gluonic corrections not accounted for in the QCD evolution of
the wave functions as well as a renormalization group transformation
from the factorization scale $\mu_F$ to the renormalization
scales. The gliding factorization scale to be used in the evolution of
the wave functions is chosen to be $\mu_F=1/\tilde{b}$ where
$\tilde b = max\{b_i\}$. 
The $b$ space form of the Sudakov factor has been calculated 
in next-to-leading-log approximation by Botts and Sterman \cite{BLS}.\\ 
As mentioned before, exclusive charmonium
decays have beeen analysed several times before, e.g.\
\cite{dun80}. New refined analyses are however necessary
for several reasons: in previous analyses the standard hard scattering
approach has been used with the
running of $\als$ and the evolution of the wave function ignored. In
the case of the $\chi_{\c J}$ also the colour-octet contributions have
been disregarded. Another very important disadvantage of some of the 
previous analyses is the use of light hadron distribution amplitudes (DAs), 
representing wave functions integrated over transverse momenta, 
that are strongly concentrated in the end-point regions. 
Despite of their frequent use, such \das\ were always subject to
severe criticism, see e.g.\ \cite{isg}. In the case of the pion,
they lead to clear contradictions to the recent CLEO 
data \cite{CLEO} on the $\pi\gamma$ transition form factor, 
$F_{\pi\gamma}$. As detailed analyses unveiled, the $F_{\pi\gamma}$
data require a \da{} that is close to the asymptotic form $\propto x(1-x)$ 
\cite{Rau96,mus97}. Therefore, these end-point region
concentrated \das\ should be discarded for the pion and perhaps also
for other hadrons like the nucleon \cite{BK96}. 

\section{RESULTS FOR $\jp\;(\psi')\to\BbB$}
According to the statements put forward in Sects.\ 1 and 2, higher
Fock-state contributions are neglected in this case. \\
The $\jp$ colour-singlet component is written in a covariant fashion 
\bea
  |\jp; \,q,\lambda;\c\cbar_1\rangle &=&\, \frac{\delta_{ab}}{\sqrt{3}}\,
      \left (\frac{f_{\jp}}{2\,\sqrt{6}} \right )\nn\\
      &\times&  \frac{1}{\sqrt{2}} (\qsla+\mjp) \esla(\lambda)\, .
  \label{Jpsistate}
\eea
The $\jp$ decay constant $f_{\jp}$ ($=f^{(1)}(^3S_1)$) is obtained
from the electronic $\jp$ decay width and found to be 409 MeV. Except 
in phase space factors, the baryon masses are ignored and $\mjp$ is 
replaced by $2\mc$ for consistency since the binding energy is an 
$O(v^2)$ effect.\\
From the permutation symmetry between the two $u$ quarks and from the 
requirement that the three quarks have to be coupled in an isospin 
$1/2$ state it follows that there is only one independent scalar wave 
function in the case of the nucleon if the 3-component of the orbital 
angular momentum is assumed to be zero. Since $\su3$ is a good
symmetry, only mildly broken by quark mass effects, one may also
assume that the other octet baryons are described by one scalar 
wave function. It is parameterized as
\be
  \Psi^{B_8} _{123}(x,{\bf k_{\perp}}) =
  \frac{f_{B_8}}{8\sqrt{3!}}\,
   \phi^{B_8}_{123}(x)\,
           \Omega_{B_8} (x,{\bf k_{\perp}}) 
\label{Psiansatz}
\ee
in the transverse momentum space. The set of indices 123 refers to the quark
configuration $\u_+\,\u_-\,\d_+$; the wave functions for other quark
configurations are to be obtained from (\ref{Psiansatz}) by
permutations of the indices. The transverse momentum dependent part
$\Omega$ is parameterized as a Gaussian in $k^2_{\perp i}/x_i$
($i=1,2,3$). The transverse size parameter, $a_{B_8}$, appearing in that
Gaussian, as well as the octet-baryon decay constant, $f_{B_8}$, are
assumed to have the same value for all octet baryons. In \cite{BK96}
these two parameters as well as the nucleon \da{} have been determined
from an analysis of the nucleon form factors and valence quark structure
functions at large momentum transfer ($a_{B_8}=0.75 \gev^{-1}$;
$f_{B_8}=6.64\times 10^{-3} \gev^2$ at a scale of reference $\mu_0=1
\gev$). The \da{} has been found to have the simple form
\be
  \phi^N_{123} (x,\mu_0) =  60x_1 x_2 x_3 \, \left[ 1 + 3 x_1 \right]. 
  \label{phiFIT}
\ee
It behaves similar to the asymptotic form, only the position of the
maximum is shifted slightly.
For the hyperon and decuplet baryon \das\ suitable
generalizations of the nucleon \da\ are used.\\
Calculating the hard scattering amplitude from the Feynman graphs for 
the elementary process $\c\cbar\to 3\g^*\to 3(\q\qbar)$ and working out the
convolution (\ref{structure}), one obtains the 
widths for the $\jp$ decays into $\BbB$ pairs listed and compared to
experimental data in Table \ref{tab:1S}.
%
\begin{table*}
  \begin{center}
  \begin{tabular}{|c||c|c|c|c|c|c|} \hline
    \rule{0cm}{8mm} channel & $p\overline p\;\;$ &
    $\Sigma^0\overline{\Sigma}{}^0\;\;$ & $\Lambda\overline \Lambda\;\;$ & 
    $\Xi^-\overline{\Xi}{}^+\;\;$ & $\Delta^{++}\overline{\Delta}{}^{--}\;\;$ &
    $\Sigma^{*-}\overline{\Sigma}{}^{*+}$ \\ \hline\hline
    $\Gamma_{3g}$  & 174 & 113 & 117 & 62.5 & 65.1 & 40.8\\ \hline  
    $\Gamma_{\rm exp}$ \cite{PDG} & $188\pm 14$ 
            & $110\pm 15$ &  $117\pm 14$ & $78\pm 18$ & $96\pm 26$ & $45\pm 6$  
            \\ \hline
  \end{tabular}
  \end{center}
  \caption[]{Results for $\jp\to\BbB$ decay widths (in [eV]) taken
    from \cite{bol97}. The three-gluon contributions are evaluated
    with $\mc=1.5$ GeV, $\LQCD=210$ MeV and 
    $a_{B_{10}}=0.85\,\gev^{-1}$.
\label{tab:1S}}
\end{table*}
As can be seen from that table a rather good agreement with the data
is obtained.\\
In addition to the three-gluon contribution there is also an
isospin-violating electromagnetic one generated by the subprocess
$\c\cbar\to\gamma^*\to 3(\q\qbar)$. According to \cite{bol97}
this contribution seems to be small.\\
The extension of the perturbative approach to the baryonic decays of
the $\psi'$ is now a straightforward matter. One simply has to
rescale the $\jp$ widths by the ratio of the corresponding electronic widths
\bea
  \label{sca}
\Gamma(\psi'\to\BbB)\,=\,
         \frac{\rho_{\mathrm p.s.}(m_{B}/M_{\psi'})}
                                {\rho_{\mathrm p.s.}(m_{B}/\mjp)}
                               \hspace{1cm}\nn\\
    \hspace{1cm}\times \frac{\Gamma(\psi'\to e^+ e^-)}{\Gamma(\jp \to e^+ e^-)}\;
                   \Gamma(\jp\to\BbB)\, ,     
\eea
where $\rho_{\mathrm p.s.}(z)=\sqrt{1-4z^2}$ is the phase space
factor. Contrary to previous calculations the $\psi'$
and the $\jp$ widths do not scale as $(\mjp/M_{\psi'})^8$ since the 
hard scattering amplitude is evaluated with $2\mc$ instead with the 
bound-state mass. Results for the baronic decay widths of
the $\psi'$ are presented in Table \ref{tab:2S}.
%
%
\begin{table*}
  \begin{center}
  \begin{tabular}{|c||c|c|c|c|c|c|} \hline
    \rule{0cm}{8mm} channel & $p\overline p\;\;$ &
    $\Sigma^0\overline{\Sigma}{}^0\;\;$ & $\Lambda\overline \Lambda\;\;$ & 
    $\Xi^-\overline{\Xi}{}^+\;\;$ & $\Delta^{++}\overline{\Delta}{}^{--}\;\;$ &
    $\Sigma^{*-}\overline{\Sigma}{}^{*+}$ \\ \hline\hline
    $\Gamma_{3g}$  & 76.8 & 55.0 & 54.6 & 33.9 & 32.1 & 24.4\\ \hline  
    $\Gamma_{\rm exp}$ \cite{BESC} & $76\pm 14$ 
            & $26\pm 14$ &  $58\pm 12$ & $23\pm 9$ & $25\pm 8$ & $16\pm 8$ \\ 
    {\phantom{$\Gamma_{exp}$}} \cite{PDG}  & $53\pm 15$ 
            & &   &  &  & \\ \hline
  \end{tabular}
  \end{center}
  \vspace*{-0.3cm} 
 \caption[]{The three-gluon contributions to the $\psi'\to \BbB$ 
    decay widths (in [eV]) taken from \cite{bol97}.
\label{tab:2S}}
\end{table*}
Again good agreement between theory and experiment is to be seen with
perhaps the exception of the $\Sigma^0\overline{\Sigma}{}^0\;\;$
channel. An additional factor of $(\mjp/M_{\psi'})^8$ ($\approx 0.25$)
in (\ref{sca}) would clearly lead to diagreement with the data.

\section{CONCLUSIONS}
Exclusive charmonium decays constitute an interesting laboratory 
for investigating power correstions and higher Fock-state contributions. 
In particular one can show that in the decays of the $\chi_{\c J}$
the contributions from the next-higher charmonium Fock state,
$\c\cbar\g$, are not suppressed by powers of $\mc$ or $v$ as compared
to the $\c\cbar$ Fock state and therefore have to be included
for a consistent analysis of these decays. For $\jp$ ($\psi'$) decays
into $\BbB$ pairs the situation is different: Higher Fock-state
contributions are suppressed by powers of $1/\mc$ and $v$. Indeed, as
an explicit analysis reveals, with plausible baryon wave functions a
reasonable description of the baryonic $\jp$ ($\psi'$) decay widths
can be obtained alone from the $\c\cbar$ Fock state.

\end{document}